\documentclass[%
reprint,nofootinbib,
superscriptaddress,
 amsmath,amssymb,
 aps,
floatfix,
]{revtex4-2}

\usepackage{graphicx}
\usepackage{dcolumn}
\usepackage{bm}
\usepackage{float}
\usepackage{booktabs}
\usepackage{setspace}
\usepackage{caption}

\newcommand{\beq}{\begin{equation}}
\newcommand{\eeq}{\end{equation}}
\newcommand{\bea}{\begin{eqnarray}}
\newcommand{\eea}{\end{eqnarray}}
\newcommand{\beqn}{\begin{equation*}}
\newcommand{\eeqn}{\end{equation*}}
\newcommand{\bean}{\begin{eqnarray*}}
\newcommand{\eean}{\end{eqnarray*}}

\DeclareMathOperator{\sech}{sech}

\begin{document}

\title{New generalization of the simplest $\alpha$-attractor $T$ model}

\author{Gabriel Germ\'an}
\email{e-mail: gabriel@icf.unam.mx}
\affiliation{Instituto de Ciencias F\'{\i}sicas, Universidad Nacional Aut\'onoma de M\'exico, Avenida Universidad s/n, Cuernavaca, Morelos 62210, Mexico}

\date{\today}

\begin{abstract}
The simplest $\alpha$-attractor $T$ model is given by the potential $V=V_0\tanh^2(\lambda\phi/M_{pl})$. However its generalization to the class of models of the type $V = V_0 \tanh^p (\lambda \phi/M_{pl})$ is difficult to interpret as a model of inflation for most values of $p$. Keeping the basic model, we propose a new generalization, where the final potential is of the form $V = V_0 (1-\sech^p (\lambda \phi/M_{pl}))$, which does not present any of the problems that plague the original generalization, allowing a successful interpretation as a model of inflation for any value of $p$ and, at the same time, providing the potential with a region where reheating can occur for any $p$ (including odd and fractional values) without difficulty. In the cases $p = 1, 2, 4$ we obtain the solutions $r(n_s, N_{ke})$ where $r$ is the tensor-to-scalar ratio, $n_s$ the spectral index and $N_{ke}$ the number of $e$-folds during inflation. We also show how these solutions connect to the $\phi^2$ monomial. 
\end{abstract}

\maketitle
\section {\bf Introduction}\label{INT}

Inflation models of the type $\alpha$-attractors have captivated considerable attention in recent years because they cover a substantial part of the observationally favorable region reported mainly by the Planck Collaboration \cite{Akrami:2018odb}  (see \cite{Kallosh:2013hoa}-\cite{Kallosh:2019hzo} for most of the basic material and \cite{Odintsov:2016vzz}-\cite{Shojaee:2021zmf} for a sample of subsequent work on the subject). These are well-motivated models and their origin is traced to conformal, superconformal, and supergravity theories all of which are well grounded mathematically. The simplest model of $\alpha$-attractors is given by a potential of the form
\begin{equation}
V=V_0\tanh^2(\lambda\frac{\phi}{M_{pl}})\, ,
\label{tanhbasic} 
\end{equation}
where $\lambda$ ($\lambda =1/\sqrt{6\alpha}$ in the original notation) is directly related to the curvature of the inflaton scalar manifold and $M_{pl}$ is the reduced Planck mass $M_{pl}=2.44\times 10^{18} \,\mathrm{GeV}$ however, in the plots, we work in Planck units such that $M_{pl}=1$. This basic model is generalized to the class of models $V=V_0\tanh^p(\lambda\phi/M_{pl})$ characterized by the parameter $p$  \cite{Kallosh:2013yoa}. However, values of $p$ other than $p=2$ present certain difficulties in being interpreted as inflation models.

In this article we have tried a different generalization from the previous one but at the same time keeping the basic structure that seems so promising. Therefore we generalize the basic potential $V =V_0\tanh^2(\lambda\phi/M_{pl})=V_0(1-\sech^2(\lambda\phi/M_{pl}))$ to the form $V=V_0(1-\sech^p(\lambda\phi/M_{pl}))$. This small modification brings with it important changes in the class of resulting models. The models being well defined for every reasonable value of $p$, allowing a region where reheating can occur for any $p$, including odd and fractional values, and covering practically the entire phenomenologically acceptable region in the $n_s$~vs.~$r$ plane.

The organization of the article is as follows: In Sec. \ref{THE} we discuss the new generalization of the basic model that we propose and show how the resulting potential is positive definite for every value of the power $p$. We also discuss the expansion of the model around its minimum. In particular, we see that the dependency on $p$ is weak, being only a multiplicative constant of the leading quadratic term, while the dependency of the previous generalized model is very strong, being a power of the inflaton. This has the consequence that the new generalization allows reheating for any $p$  (including odd and fractional values). In Sec. \ref{PRO}, given the impossibility of carrying out an analytical study for arbitrary $p$, we consider several interesting examples with $p = 1, 2, 4$ and we write the potential in terms of the observables $n_s$ and $r$. This allows us to obtain the limit of the potential when $n_s\rightarrow 1-r / 4$, equivalently $\lambda\rightarrow 0$, showing that the potential is reduced to the quadratic potential $V=\frac{1}{2}m^2\phi^2$ connecting the solution,  in the $n_s$~vs~$r$ plane, with the monomial $\phi^2$. We show figures for the number of $e$-folds during inflation $N_ {ke}$, the tensor-to-scalar ratio  $r$ and the inflation scale as functions of $p$ for various values of the parameter $\lambda$. Fnally, Sec. \ref{CON} contains our conclusions on the main points discussed in the article.
\section {\bf The model}\label{THE}

\begin{figure}[tb]
\begin{center}
\captionsetup{format=plain,justification=centerlast}
\includegraphics[trim = 0mm  0mm 1mm 1mm, clip, width=8.5cm, height=6.cm]{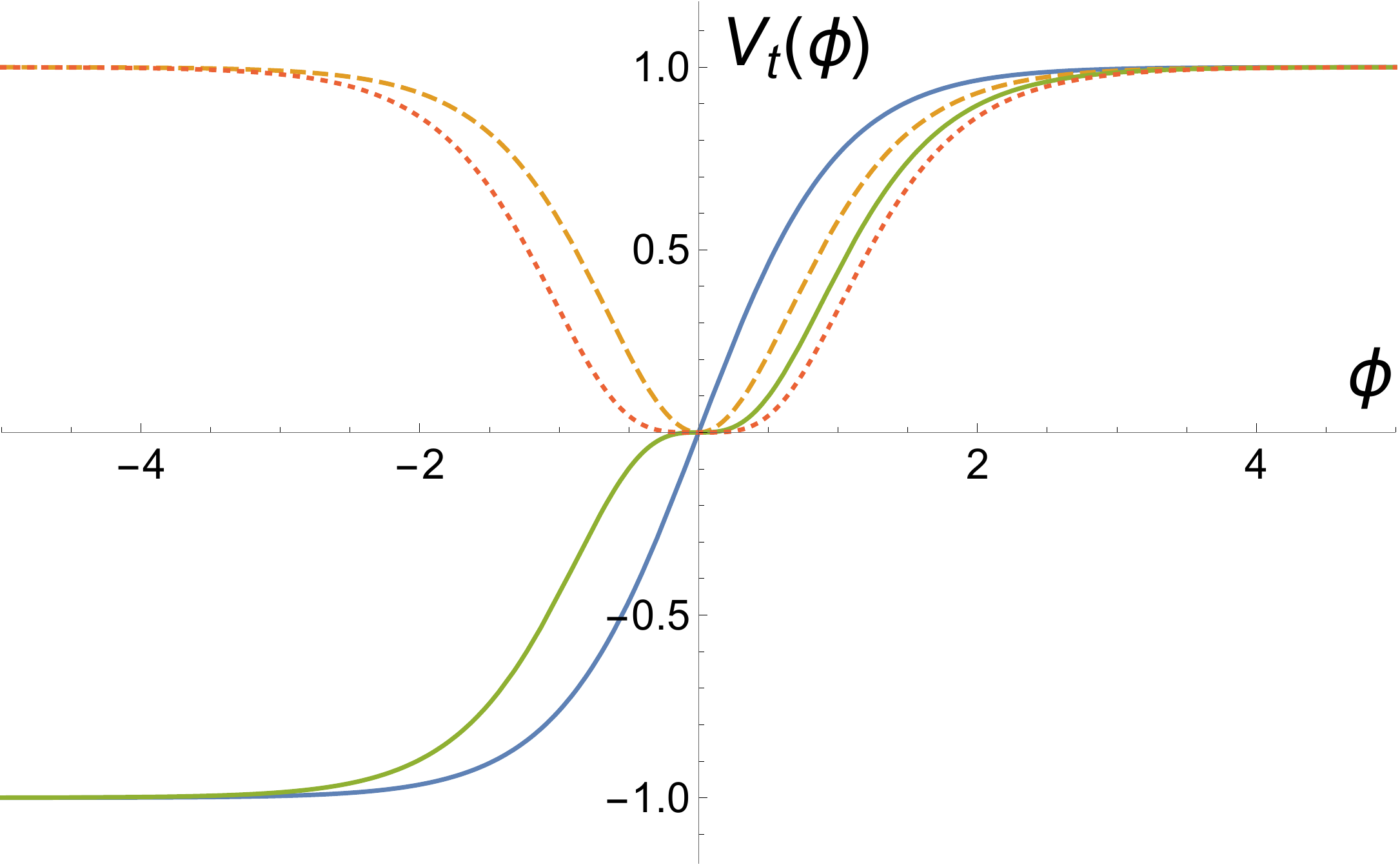}
\includegraphics[trim = 0mm  0mm 1mm 1mm, clip, width=8.5cm, height=6.cm]{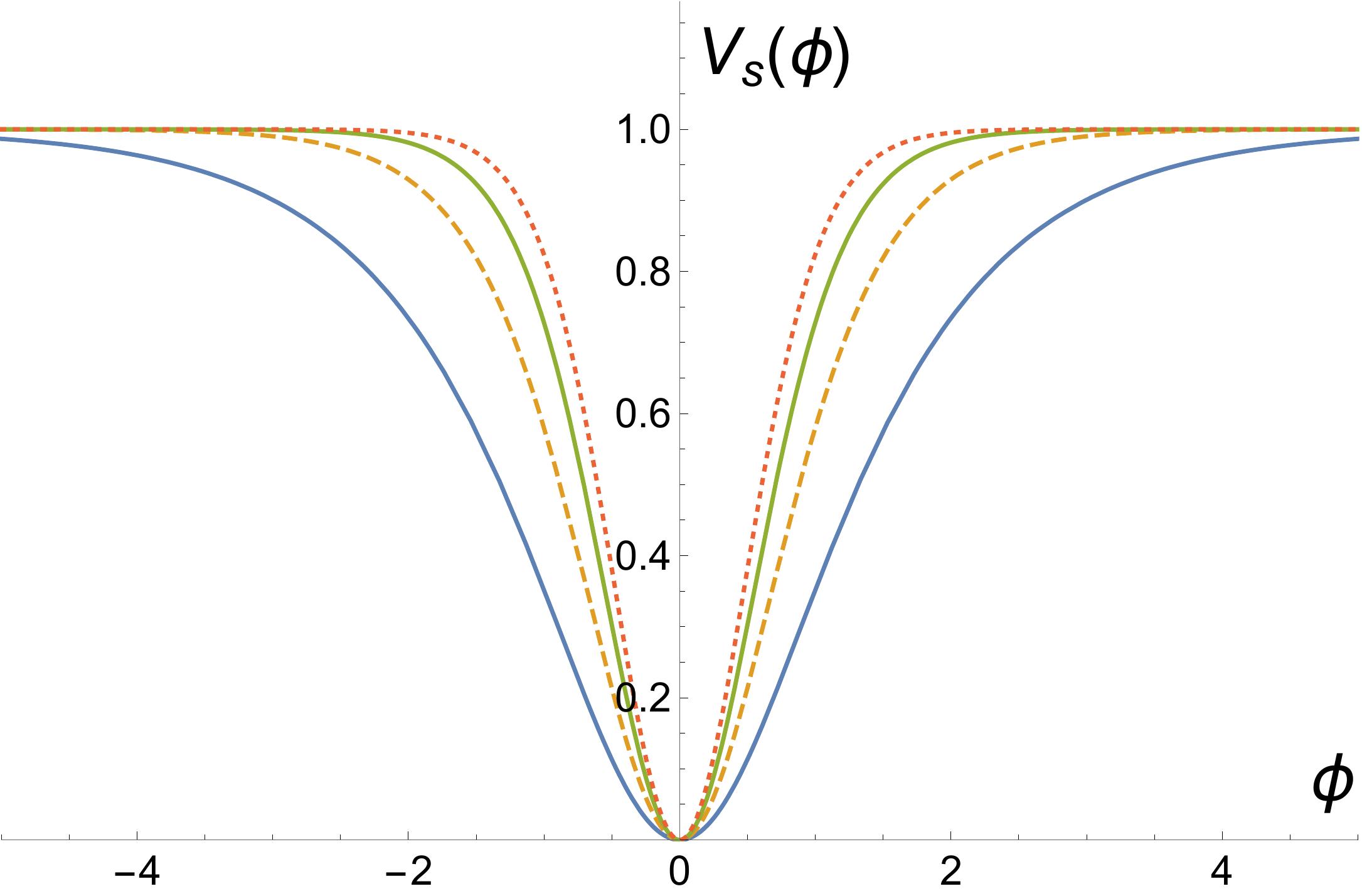}
\caption{\small  Looking at the first quadrant from left to right $p=1,2,3,4$ for the upper figure, while  $p= 4,3,2,1$ for the lower figure. Odd-$p$ cases are difficult to interpret as inflation models for the $\tanh^p$ potential  given by Eq.~\eqref{potanh}. This is not the case for the modified $\sech^p$ potential, Eq.~\eqref{potsech} which is viable even for fractional values of the parameter $p$. Another striking feature of the $\sech^p$  potential is shown in Fig.\,\ref{potanwide} below for the even-$p$ cases.
}
\label{potan}
\end{center}
\end{figure}

\begin{figure}[tb]
\begin{center}
\captionsetup{format=plain,justification=centerlast}
\includegraphics[trim = 0mm  0mm 1mm 1mm, clip, width=8.5cm, height=6.cm]{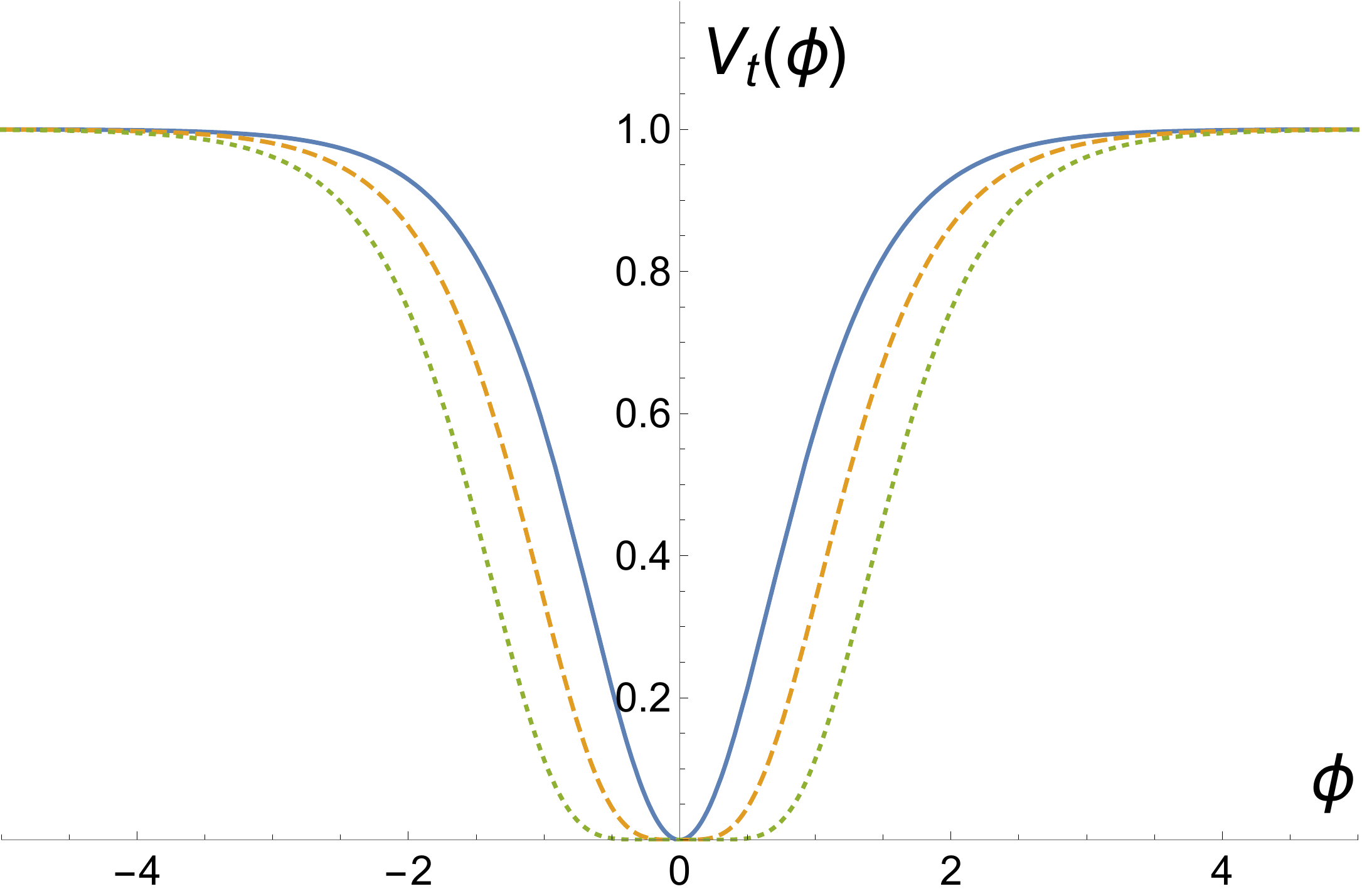}
\includegraphics[trim = 0mm  0mm 1mm 1mm, clip, width=8.5cm, height=6.cm]{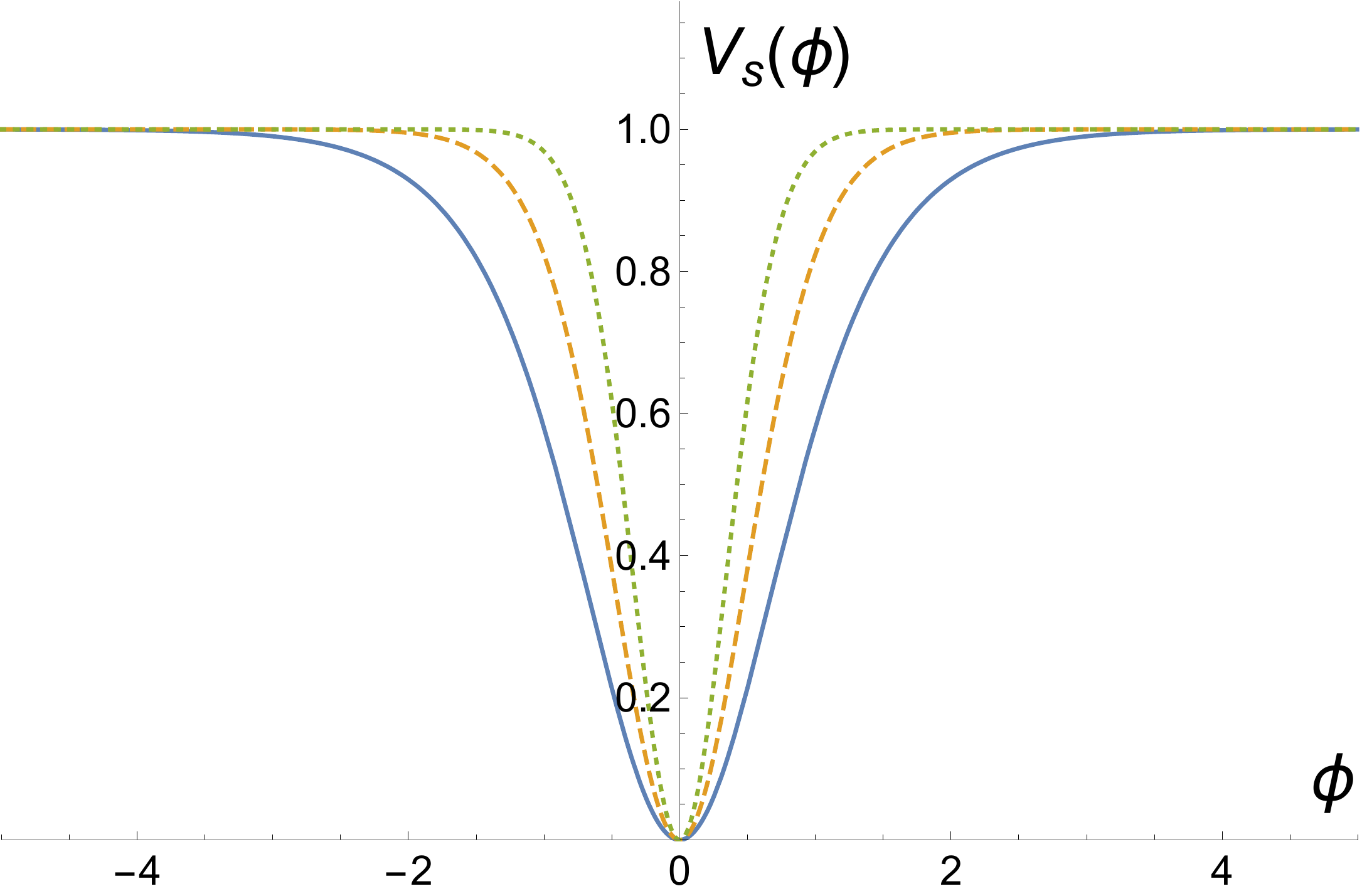}
\caption{\small Looking at the first quadrant from left to right $p=2, 4, 8$ for the upper figure, while  $p= 8, 4, 2$ for the lower figure. Around the minimum the $\tanh^p$ potential,  given by Eq.~\eqref{potanh}, is flatter than the $\sech^p$ potential, Eq.~\eqref{potsech}, for $p>2$. This is easily understood because the $\tanh^p$ has a strong $p$-dependence at the origin with the leading term going like $(\lambda \phi)^p$ while $p$ in the $\sech^p$ potential is only a proportionality constant of the leading $(\lambda \phi)^2$ term. Thus, the $\sech^p$ potential  is quadratic at the origin for $any$ $p$ (see Eqs.~\eqref{potorigint} and \eqref{potorigins}). This suggest that we can have very good inflationary properties while viable reheating for a wide range of $p$ values, including odd and fractional ones.}
\label{potanwide}
\end{center}
\end{figure}

Without going into the details of the construction of the $\alpha$-attractor models we begin by writing \cite{Akrami:2017cir}
\begin{equation}
\frac{1}{\sqrt{-g}}\mathcal{L}=\frac{1}{2}M_{pl}^2R-\frac{1}{2}M_{pl}^2\frac{(\partial_{\mu}\Psi)^2}{(1-\lambda^2\Psi^2)^2} -V(\Psi),
\label{Lagrangian} 
\end{equation}
as a phenomenological model and propose a function for $V(\Psi)$ where $\Psi \propto \tanh(\lambda \phi/M_{pl})$ makes
$\phi$ a canonically normalized field identified with the inflaton. The simplest $\alpha$-attractor model is given by  \cite{Kallosh:2013yoa}
\begin{equation}
V(\Psi) \propto \Psi^2 ,
\label{Ftan} 
\end{equation}
generalized as a simple power of $\Psi$ to
\begin{equation}
V(\Psi) \propto \Psi^{p} ,
\label{FtanG} 
\end{equation}
in such a way that the potential for the inflaton can be written in the form
\begin{equation}
V_t=V_0\tanh^p(\lambda\frac{\phi}{M_{pl}})\,.
\label{potanh} 
\end{equation}
As stated before, the $p=2$ case gives the simplest $\alpha$-attractor model. However more general cases of the parameter $p$ are difficult to interpret as models of inflation giving rise to unattractive potentials (see Fig.\,\ref{potan}). Thus, we would like to keep the very nice features of the $\tanh^2$ potential while at the same time generalize the model to well defined and viable potentials. 
We could try a close expression to the one before by noticing that $1-\Psi^2 \propto 1-\tanh^2(\lambda \phi/M_{pl})=\sech^2(\lambda \phi/M_{pl})$. Thus, we propose the following function
\begin{equation}
V(\Psi) \propto 1-(1-\Psi^2) ,
\label{Fsec} 
\end{equation}
which is, of course, exactly the same simple function as in \eqref{Ftan}  but written in a suggestive way. We now generalize \eqref{Fsec} to
\begin{equation}
V(\Psi) \propto 1-(1-\Psi^2)^{p/2}\,.
\label{FsecG} 
\end{equation}
Since the $\sech$ is a positive definite function, we can finally write the resulting potential as
\begin{equation}
V_s =V_0\left(1-\sech^{p}(\lambda\frac{\phi}{M_{pl}})\right)\,.
\label{potsech} 
\end{equation}
Thus, we are generalizing in a different way the $same$ basic function as before. In Fig.\,\ref{potan} we compare the potentials \eqref{potanh}  and \eqref{potsech}  for $p=1, 2, 3, 4$. We see that for $p=2$ both potentials coincide however the cases $p=1$ and $p=3$ differ markedly while the case $p=4$ is particularly different around the minimum.
The odd powers of $V_t$ as defined in Eq.~\eqref{potanh} give rise to a runaway potential featuring two plateaus at large and small values of the field. Such a potential is typically suitable for quintessential inflation, as explored in Refs. \cite{Dimopoulos:2017zvq}, \cite{Dimopoulos:2017tud} and \cite{Akrami:2017cir}.
In Fig.\,\ref{potanwide} we again compare these potentials for even values of the parameter $p$. We see that the minimum for the $\tanh^p$ potential is flatter than for the $\sech^{p}$. The $\sech^{p}$ potential is $always$ quadratic at the minimum irrespective of the value of $p$, which could even be odd or fractional. This comes about as follows, an expansion of the potential \eqref{potanh} around the minimum is 
\begin{equation}
V_t/V_0 = (\lambda \frac{\phi}{M_{pl}})^p-\frac{1}{3}p(\lambda \frac{\phi}{M_{pl}})^{p+2}+\cdot\cdot\cdot,
\label{potorigint}
\end{equation}
while the potential \eqref{potsech} behaves like
\begin{equation}
V_s/V_0 =\frac{1}{2}p (\lambda \frac{\phi}{M_{pl}})^2-\frac{1}{24}p(2+3p)(\lambda \frac{\phi}{M_{pl}})^4+\cdot\cdot\cdot.
\label{potorigins}
\end{equation}
Thus, at the minimum, we see a strong  dependence on $p$ for the $\tanh^p$ potential while for the $\sech^p$ potential $p$ is only a proportionality constant to the leading $\phi^2$-term. The higher the power $p$, the flatter the $\tanh^p$ potential. For the $\sech^p$ potential the dependence on $p$  is weak behaving as a quadratic potential for any $p$ (see Fig.\,\ref{potanwide}).  
\section {\bf Properties of the model}\label{PRO} 

We study some properties of the model defined by the Eq.~\eqref{potsech}, in particular, we eliminate the model parameters $V_0$ and $\lambda$ in terms of the observables $n_s$ and $r$ which facilitate a better understanding of the model. Typically the global scale $V_0$ is of no interest because quantities like the number of e-folds during inflation $N_{ke}$ and the observables $n_s$ and $r$ are related to the potential by ratios of the potential and its derivatives which eliminate $V_0$. For this model, however, it is not posible to solve the corresponding equations and to make some progress we are led to consider first the solution for the inflaton at horizon crossing $\phi_k$ by solving the equation for the amplitude of scalar perturbations $A_s$ 
\begin{equation}
A_s(k) =\frac{1}{24\pi ^{2}} \frac{V}{\epsilon\, M_{pl}^4}\,,
\label{IA} 
\end{equation}
which, however, involves the scale $V_0$.  The solution is given by
\begin{equation}
\sech(\lambda \frac{\phi_k}{M_{pl}})=\left(1-\frac{3A_s\pi^2r}{2V_0}M_{pl}^4\right)^{1/p}\,.
\label{sechfik} 
\end{equation}
From the equation $16\epsilon=r$ we get
\begin{equation}
\lambda=\left(\frac{r\left(1-\sech^p(\lambda \phi_k/M_{pl})\right)^2}{8p^2\sech^{2p}(\lambda \phi_k/M_{pl})(1-\sech^2(\lambda \phi_k/M_{pl}))}\right)^{1/2}\,,
\label{lambda} 
\end{equation}
where the $\sech(\lambda \phi_k/M_{pl})$ is given by Eq.~\eqref{sechfik} above. Unfortunately it is not possible to solve for $V_0$ for a general $p$ thus, in what follows, we discuss a few particular cases. 

\subsection {\bf The $p=1$ case}\label{caso1} 
\begin{figure*}
\captionsetup{format=plain,justification=centerlast}
\begin{center}$
\begin{array}{ccc}
\includegraphics[width=3.25in]{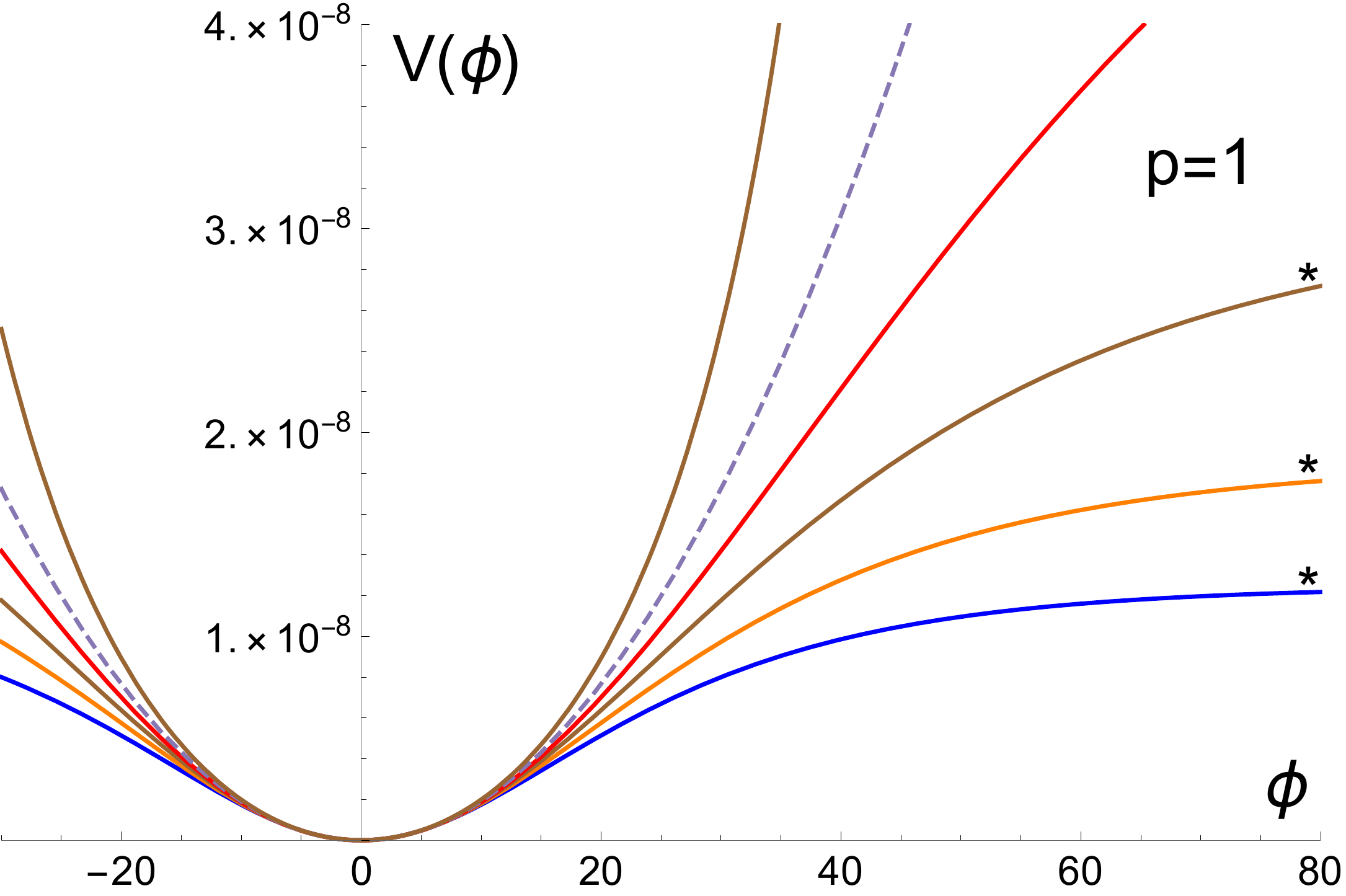}&
\includegraphics[width=3.25in]{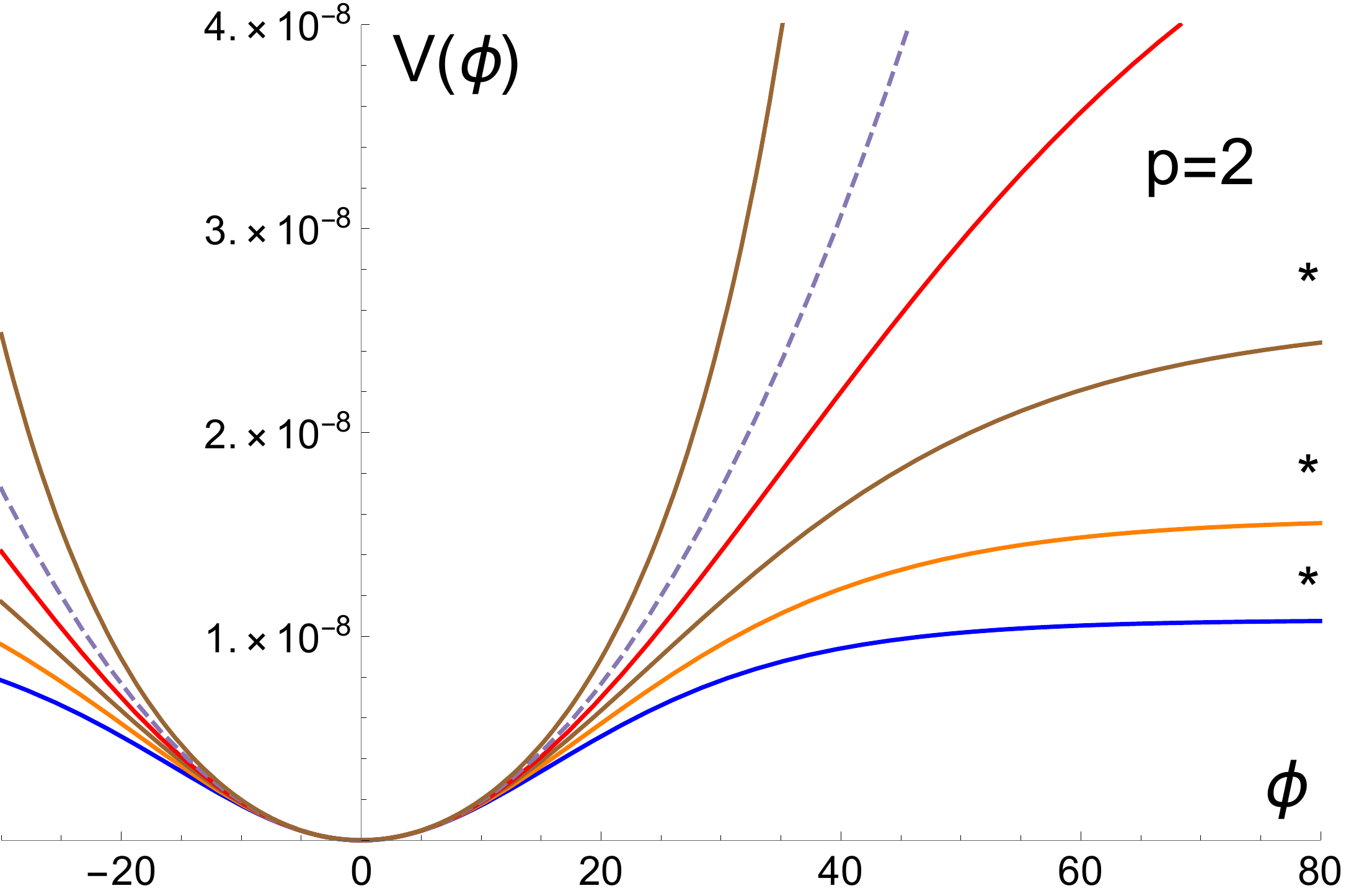}\\
\includegraphics[width=3.25in]{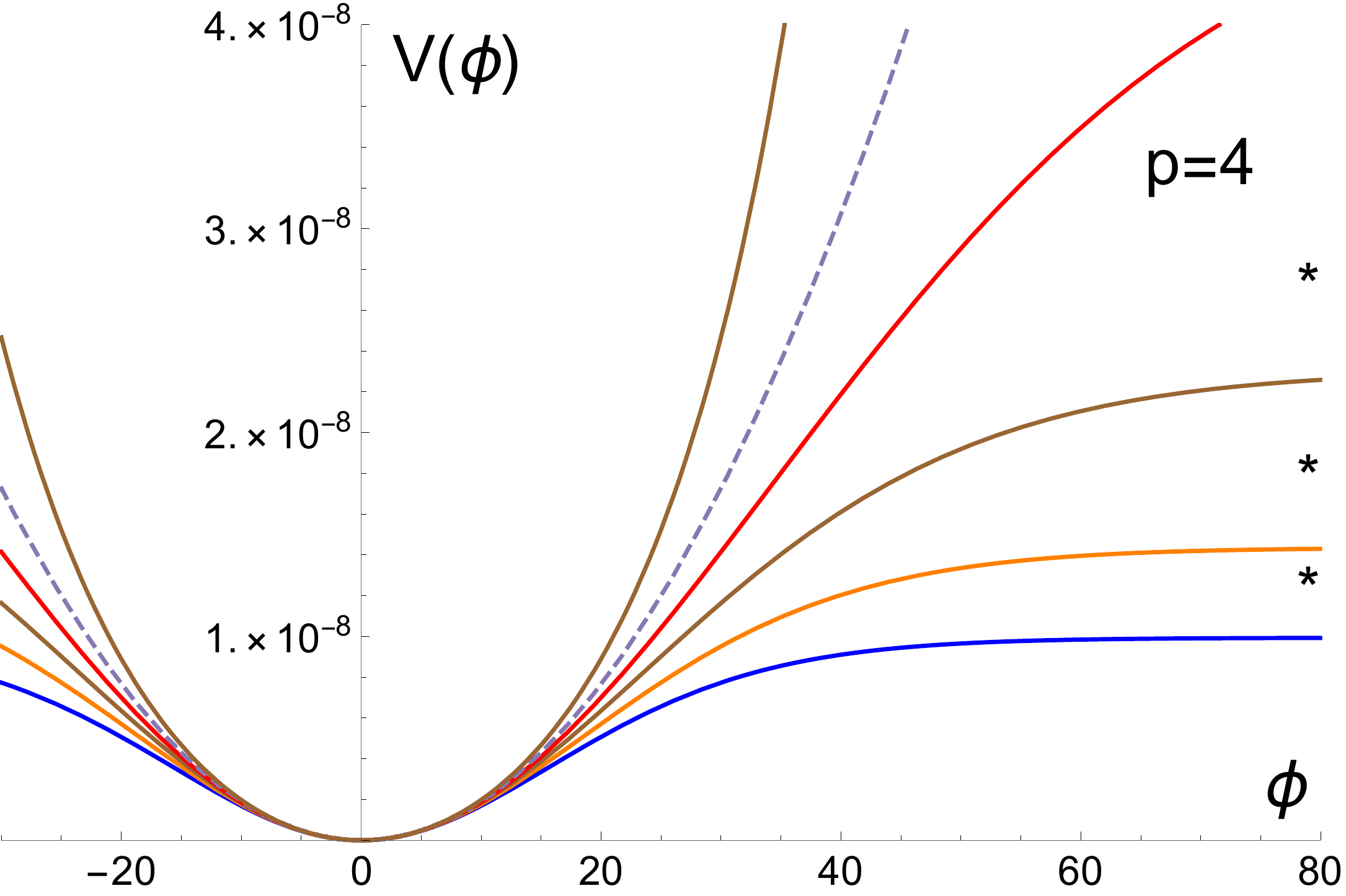}&
\includegraphics[width=3.25in]{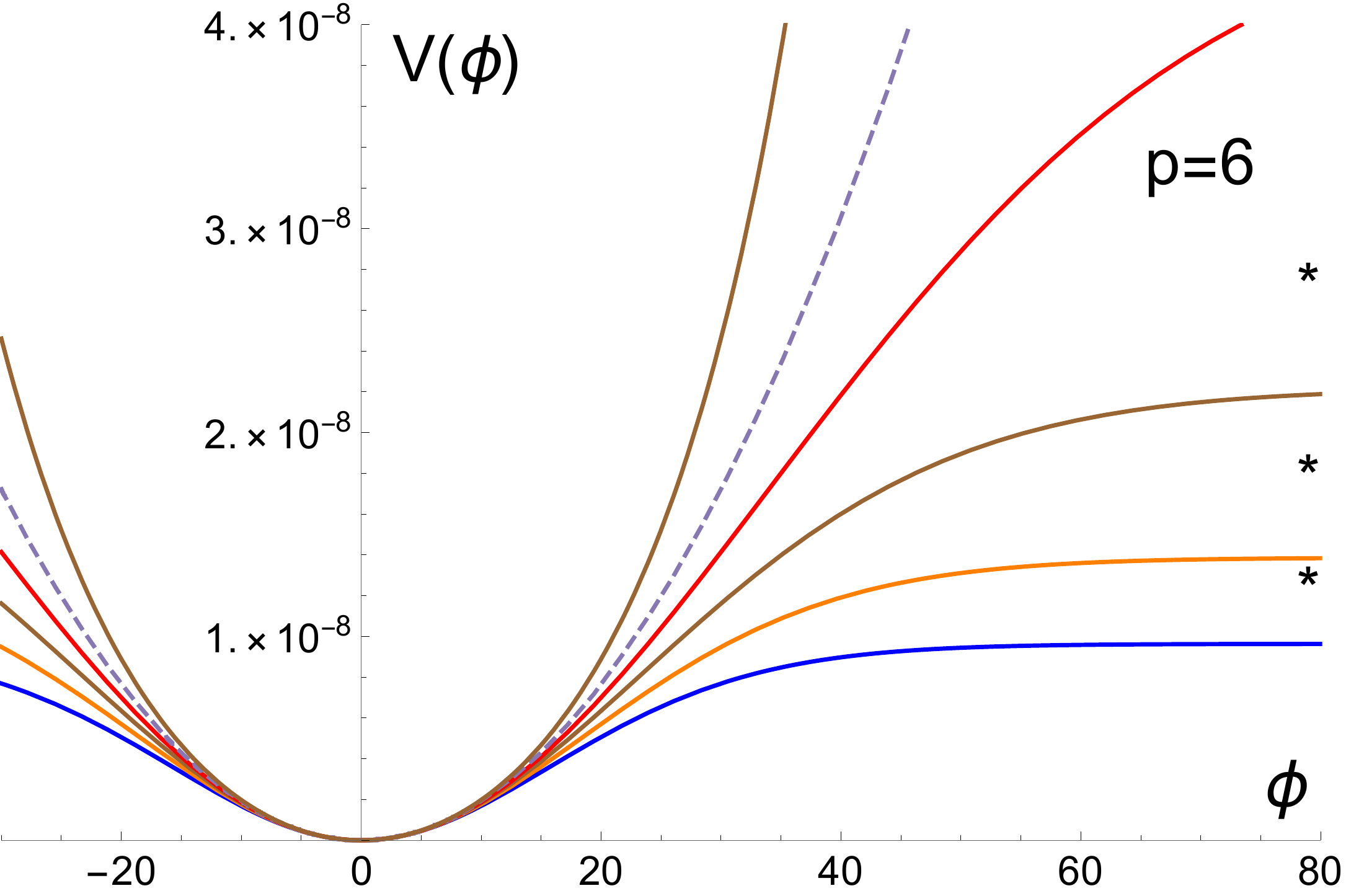} \\
\end{array}$
\end{center}
\caption{Plot of the potential $\eqref{potsech}$ for $p=1, 2, 4, 6$ for various values of $r$ reaching $r=4\delta_{n_s}$ (dashed curve) where the potential transitions to a monomial $V =\frac{3}{64}M_{pl}^2 A_s\pi^2r^2\phi^2$ (see specific examples in Section \ref{PRO}). As $p$ grows the potential flattens (the asterisks on the rhs have exactly the same coordinates and are there for reference only).  The curve past the dashed curve indicates that the potential further transitions to a $\sec^p$ potential in a range of values for $r$ which is, however, phenomenologically unacceptable. The values of $r$ used in the plots are (counterclockwise starting with the flatter curve) $r=0.1, 0.11, 0.12, 0.13, 0.1404$ (dashed) and 0.16. Smaller, phenomenological values of $r$ were not used in order to show the transition which occurs for large $r$. }
\label{pi}
\end{figure*}
\begin{figure}[tb]
\begin{center}
\captionsetup{format=plain,justification=centerlast}
\includegraphics[width=8.5cm]{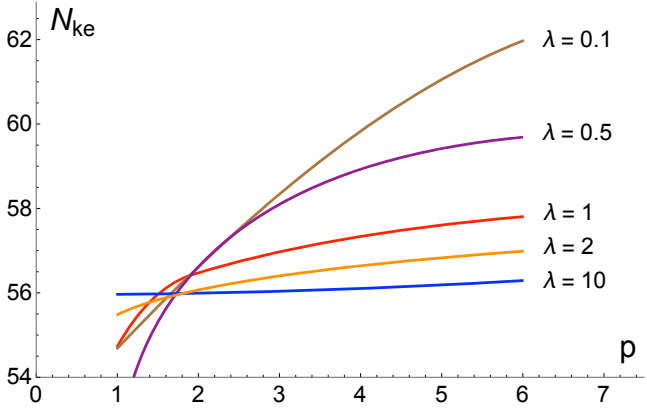}
\caption{\small The number of $e$-folds $N_{ke}$ from $\phi_k$ at horizon crossing to the end of inflation at $\phi_e$ as a function of $p$ for various values of the parameter $\lambda$. The lowest value of the sample corresponds to $N_{ke}\approx 52$ for $\lambda=0.5$ (not shown) while the highest value $N_{ke}\approx 62$ occurs for $\lambda= 0.1$.
}
\label{Nkep}
\end{center}
\end{figure}
\begin{figure}[tb]
\begin{center}
\captionsetup{format=plain,justification=centerlast}
\includegraphics[trim = 0mm  0mm 1mm 1mm, clip, width=8.5cm, height=6.cm]{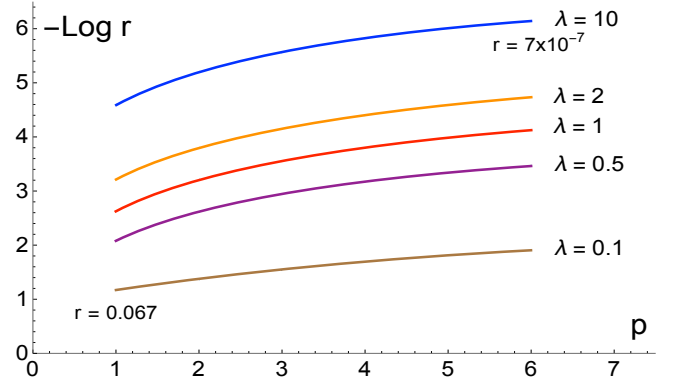}
\includegraphics[trim = 0mm  0mm 1mm 1mm, clip, width=8.5cm, height=6.cm]{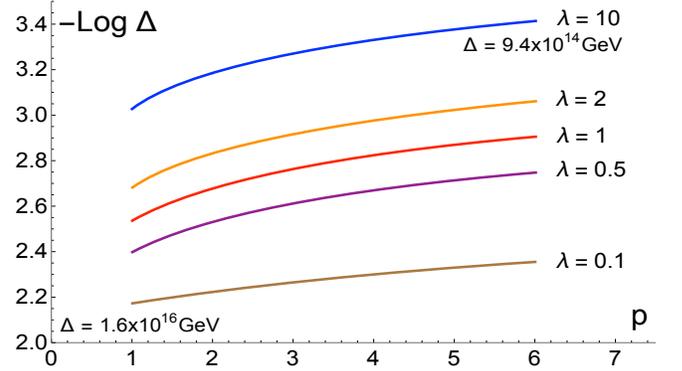}
\caption{\small The upper figure shows (minus) the logarithm of the tensor-to-scalar ratio $r$ as a function of the power $p$ for various values of $\lambda$. The small numbers in the corners indicate the lower and maximum values of $r$ for the drawn curves corresponding e.g., $r=0.067$ to the line $\lambda=0.1$ and power $p=1$. The lower figure is a similar plot for (minus) the logarithm of the scale of inflation $\Delta=V_k^{1/4}$ in Planck units, although the reference numbers are already dimensionful. 
}
\label{rp}
\end{center}
\end{figure}
\begin{figure}[tb]
\begin{center}
\captionsetup{format=plain,justification=centerlast}
\includegraphics[width=8.5cm]{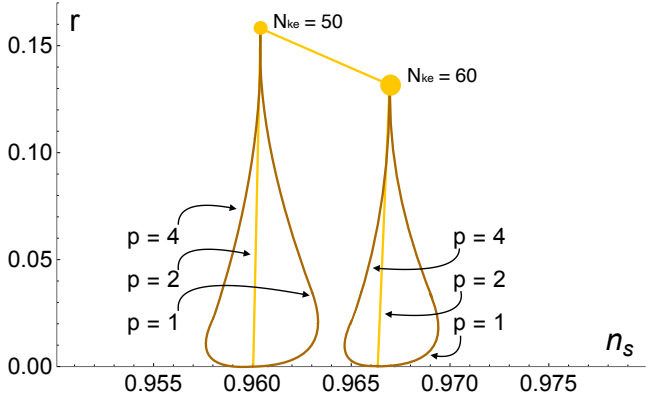}
\caption{\small  Plot of the  tensor-to-scalar ratio $r$ as a function of the spectral index $n_s$ for the models defined by $p=1, 2, 4$. This is shown for a number of $e$-folds during inflation $N_{ke}=50$ and $N_{ke}=60$. The region inside the ``bags" gets filled as $p$ continuously sweeps from $p=1$ to $p=4$ covering most of the interesting zone bounded by the results presented by Planck's  article  \cite{Akrami:2018odb}.}
\label{rdens}
\end{center}
\end{figure}
\begin{figure}[t!]
\captionsetup{format=plain,justification=centerlast}
\begin{center}
\includegraphics[trim = 0mm  0mm 1mm 1mm, clip, width=8.5cm, height=6.5cm]{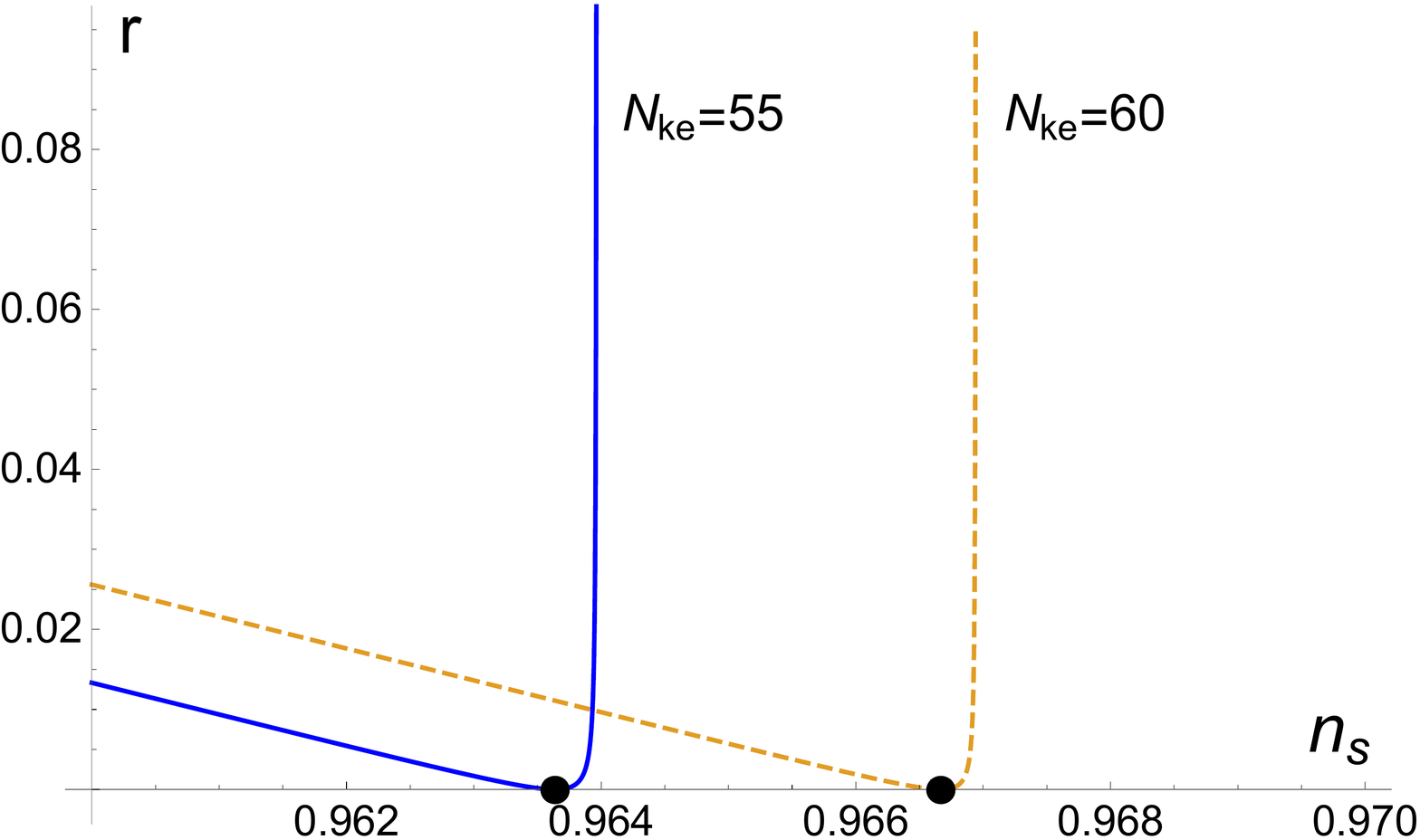}
\includegraphics[trim = 0mm  0mm 1mm 1mm, clip, width=8.5cm, height=7.cm]{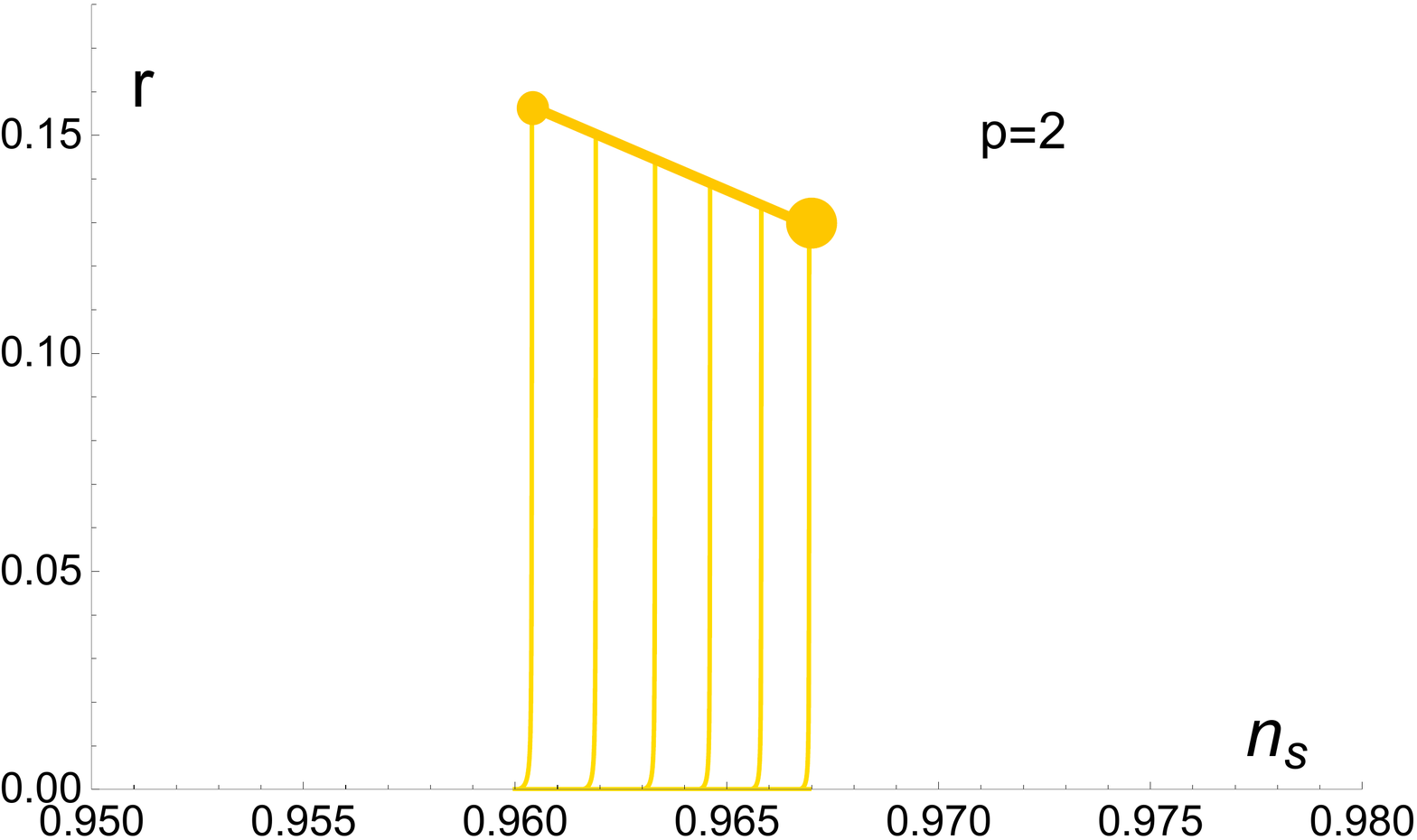}
\end{center}
\caption{The upper figure shows $r(n_s)$ for $N_{ke}=55, 60$ according to Eq.~\eqref{rdensNkep2} with only the condition on the denominator $1+N_{ke} (N_{ke} \delta_{n_s}-2)>0$ to guarantee a positive $r$. The dots occur when $N_{ke} \delta_{n_s}-2=0$ (thus, $r=0$) and are there for reference only. For $N_{ke} \delta_{n_s}-2<0$ we get the almost vertical rhs branch of the solution with $n_s$ barely increasing from the dot. For $N_{ke} \delta_{n_s}-2>0$ we get the lhs branch with $n_s$ decreasing from the dot. Substituting $r$, as given by Eq.~\eqref{rdensNkep2}, back into Eq.~\eqref{Nkexplicit} we find that to get $N_{ke}$,  $N_{ke} \delta_{n_s}-2$ has to be negative within Planck's ranges for $n_s$ and $r$ \cite{Akrami:2018odb}. From here we conclude that the lhs branch is unphysical. In the lower figure we plot $r(n_s)$ supplemented with the conditions \eqref{conditions} which guarantee a consistent solution for $r$ for (from left to right) $N_{ke} =50, 52, 54, 56, 58, 60$. }
\label{branch}
\end{figure}
The $p=1$ case corresponds to the model \cite{Pal:2009sd}, \cite{Pal:2010eb} (see also \cite{German:2020eyq},\cite{German:2020cbw})
\begin{equation}
V=V_0\left(1-\sech(\lambda\, \frac{\phi}{M_{pl}})\right)\,.
\label{potsech1} 
\end{equation}
From the equation
\begin{equation}
n_{s} =1+2\eta -6\epsilon ,  
\label{Ins} \\
\end{equation}
written in the form $\delta_{n_s}+2\eta-6\epsilon=0$, where $\delta_{n_s}$ is defined as $\delta_{n_s}\equiv 1-n_s$, we obtain
\begin{equation}
V_0 =\frac{3A_s\pi^2r(24\delta_{n_s}-r+\sqrt{17r^2+16r\delta_{n_s}+64\delta_{n_s}^2})}{16(4\delta_{n_s}-r)}M_{pl}^4 ,  
\label{V01} \\
\end{equation}
in this case the potential \eqref{potsech1} can be written in terms of the observables $n_s$ and $r$ as follows
\begin{widetext}
\begin{equation}
V =\frac{3A_s\pi^2r(24\delta_{n_s}-r+R_1)}{16(4\delta_{n_s}-r)}\left(1-\sech\left(\frac{1}{8}\sqrt{45r-16\delta_{n_s}-11R_1+\frac{8\delta_{n_s}}{r}(8\delta_{n_s}+R_1)}\,\,  \frac{\phi}{M_{pl}}\right)\right)M_{pl}^4 ,  
\label{V1} \\
\end{equation}
\end{widetext}
where $R_1\equiv \sqrt{17r^2+16r\delta_{n_s}+64\delta_{n_s}^2}$
in the limit $\delta_{n_s}\rightarrow r/4$ the potential becomes
\begin{equation}
V =\frac{3}{64}M_{pl}^2 A_s\pi^2r^2\phi^2\,.
\label{Vmon1} \\
\end{equation}
This potential is exactly the potential for the monomial $V=\frac{1}{2}m^2\phi^2$ once the parameter $m$ is eliminated by means of Eq.~\eqref{IA} above. Thus, in the limit  $\delta_{n_s}\rightarrow r/4$ (equivalently $\lambda\rightarrow 0$) the potential \eqref{potsech1} transitions to the $\phi^2$ monomial as shown in Fig.\,\ref{pi}, panel $p=1$. For $r > 4\delta_{n_s}$ there is yet another transition to a $\sec(\lambda \phi/M_{pl})$ potential but we do not study it here because it is not phenomenologically acceptable with values for $r$ beyond its upper bound.  Plots for the number of $e$-folds $N_{ke}$ the tensor-to-scalar ratio $r$ and the scale of inflation $\Delta\equiv V_k^{1/4}$ for several values of the parameter $\lambda$ are given as functions of $p$ in Figs.\,\ref{Nkep} and \ref{rp}, respectively. Also, a plot of $r(n_s,N_{ke})$  in the $n_s$ versus $r$ plane for the number of e-folds $N_{ke}=50, 60$ is shown in Fig.\,\ref{rdens}, together with the $p=2$ and $p=4$ cases. From this last figure we see that the $\sech^p$ potential always ends in the $\phi^2$ monomial as expected.
\subsection {\bf The $p=2$ case}\label{caso2} 
We solve again the equation $\delta_{n_s}+2\eta-6\epsilon=0$ with the result
\begin{equation}
V_0 =\frac{6A_s\pi^2r\delta_{n_s}}{4\delta_{n_s}-r}M_{pl}^4 ,  
\label{V02} \\
\end{equation}
in this case the potential is given by
\begin{equation}
V =\frac{6A_s\pi^2r\delta_{n_s}}{4\delta_{n_s}-r}\left(1-\sech^2\left(\frac{\sqrt{\delta_{n_s}(4\delta_{n_s}-r)}}{2\sqrt{2r}}\, \frac{\phi}{M_{pl}}\right)\right)M_{pl}^4\,, 
\label{V2} \\
\end{equation}
and in the limit $\delta_{n_s}\rightarrow r/4$ the potential again transitions to
\begin{equation}
V =\frac{3}{64}M_{pl}^2 A_s\pi^2r^2\phi^2\,.
\label{Vmon2} \\
\end{equation}
In the $p=2$ case we can find a simple expression for $r$ as a function of $n_s$ and $N_{ke}$  \cite{German:2021tqs}: from the equation $\delta_{n_s}+2\eta-6\epsilon=0$ we get
\begin{equation}
\cosh^2\left(\lambda\frac{\phi_k}{M_{pl}}\right)=\frac{\delta_{n_s}+8\lambda^2+\sqrt{\delta_{n_s}^2+16\lambda^2\delta_{n_s}+64\lambda^4}}{2\delta_{n_s}},
\label{fik1a}
\end{equation}
while the solution to $\epsilon=1$ gives the end of inflation 
\begin{equation}
\cosh^2\left(\lambda\frac{\phi_{e}}{M_{pl}}\right)=\frac{1}{2}\left(1+\sqrt{1+8\lambda^2}\right)\;.
\label{fie2a}
\end{equation}
The number of e-folds $N_{ke} = -\frac{1}{M_{pl}^2}\int_{\phi_k}^{\phi_e}\frac{V}{V'}d\phi$ is
\begin{equation}
N_{ke} =\frac{1}{4\lambda^2}\left(\cosh^2\left(\lambda\frac{\phi_k}{M_{pl}}\right)-\cosh^2\left(\lambda\frac{\phi_e}{M_{pl}}\right)\right)\;,
\label{Nke}
\end{equation}
or
\begin{equation}
N_{ke} = \frac{8\delta_{n_s}-r-\sqrt{r^2+r\delta_{n_s}(4\delta_{n_s}-r)}}{\delta_{n_s}(4\delta_{n_s}-r)}\;.
\label{Nkexplicit}
\end{equation}
Solving for $r$
\begin{equation}
r = \frac{4(N_{ke}\delta_{n_s} -2)^2}{1+N_{ke} (N_{ke} \delta_{n_s}-2)}\;,
\label{rdensNkep2}
\end{equation}
This solution should be supplemented with conditions which guarantee that $r$ is a well-defined real positive number. To have a positive $r$ the condition on the denominator $1+N_{ke} (N_{ke} \delta_{n_s}-2)>0$ has to be satisfied. The numerator implies that $r=0$ if and only if $N_{ke} \delta_{n_s}-2=0$ (small dots as reference points in the upper panel of Fig.~\ref{branch}). For $N_{ke} \delta_{n_s}-2<0$ we get the almost vertical rhs branch of the solution with $n_s$ barely increasing from the dot. For $N_{ke} \delta_{n_s}-2>0$ we get the lhs branch with $n_s$ decreasing from the dot. If we substitute $r$ as given by Eq.~(26) back in the rhs of Eq.~(25) we will find that to get $N_{ke}$ (as we should)  $N_{ke} \delta_{n_s}-2$ has to be negative within Planck's ranges for $n_s$ and $r$ [1]. Thus, the lhs branch is unphysical. Also, if we combine these two conditions ($N_{ke} \delta_{n_s}-2<0$ and $1+N_{ke} (N_{ke} \delta_{n_s}-2)>0$), it is easy to show that they are equivalent to the following conditions on $n_s$
\begin{equation}
1-\frac{2}{N_{ke}} \leq n_s < 1-\frac{2}{N_{ke}}+\frac{1}{N_{ke}^2}\;.
\label{conditions}
\end{equation}
Thus, the solution given by Eq.(26) should be supplemented with the conditions (27) above. The solution (26) is plotted for various values of $N_{ke}$ in the lower panel of Fig.~\ref{branch}.
\\
\\
\subsection {\bf The $p=4$ case}\label{caso4} 
From the equation $\delta_{n_s}+2\eta-6\epsilon=0$ we obtain the result
\begin{widetext}
\begin{equation}
V_0 =\frac{3A_s\pi^2r\left(512\delta_{n_s}^2-192r\delta_{n_s}+9r^2-r\sqrt{r(256\delta_{n_s}-15r)}\right)}
{32(8\delta_{n_s}-3r)(4\delta_{n_s}-r)}M_{pl}^4 ,  
\label{V04} \\
\end{equation}
\end{widetext}
in this case the potential is given by
\begin{widetext}
\begin{equation}
V =\frac{3A_s\pi^2r R_2}
{32(8\delta_{n_s}-3r)(4\delta_{n_s}-r)} \left(1-\sech^4\left(\frac{\sqrt{2r}(8\delta_{n_s}-3r)(4\delta_{n_s}-r)}
{\sqrt{R_3^2\left(1-\sqrt{R_3/R_2}\right)}}\,  \frac{\phi}{M_{pl}}\right)\right)M_{pl}^4\,, 
\label{V4} \\
\end{equation}
\end{widetext}
where $R_2\equiv 512\delta_{n_s}^2-192r\delta_{n_s}+9r^2-r\sqrt{r(256\delta_{n_s}-15r)}$ and $R_3 \equiv 128r\delta_{n_s}-39r^2-r\sqrt{r(256\delta_{n_s}-15r)}$. In the limit $\delta_{n_s}\rightarrow r/4$ the potential becomes
\begin{equation}
V =\frac{3}{64}M_{pl}^2 A_s\pi^2r^2\phi^2\,, 
\label{Vmon4} \\
\end{equation}
as in the previous two cases. We expect this to be a general result: from Eq.~(13) we can express $r$ in terms of $\lambda$. A small $\lambda$ expansion gives
\begin{equation}
r=\frac{32 M_{pl}^2}{\phi^2}-\frac{16}{3}(2+3p)\lambda^2+\cdot\cdot\cdot,
\label{rexpan} \\
\end{equation}
or
\begin{equation}
\frac{r\phi^2}{32 M_{pl}^2}=1-\frac{\phi^2}{6M_{pl}^2}(2+3p)\lambda^2+\cdot\cdot\cdot.
\label{newrexpan} \\
\end{equation}
Thus, in the limit of vanishing $\lambda$ and using Eq.~(11), we get
\begin{equation}
V=\frac{r\phi^2}{32 M_{pl}^2}V=\frac{r\phi^2}{32 }M_{pl}^2 \frac{3}{2}A_s\pi^2r= \frac{3}{64}M_{pl}^2A_s\pi^2r^2\phi^2\,.
\label{Vproof} \\
\end{equation}
for any value of $p$.
\\
\section {\bf Conclusions}\label{CON}
Starting from the simplest monomial function for $\alpha$-attractors we have proposed a new generalization of the $T$ models leading to the potential $V = V_0 (1-\sech^p (\lambda \phi/M_{pl}))$, that does not present the difficulties of interpretation of the original generalization given by $V = V_0 \tanh^p (\lambda \phi/M_{pl})$. The resulting class of potentials have also the particularity that they are quadratic around the minimum for all values of the power $p$ giving rise to viable inflation models while at the same time presenting a region where reheating can occur without difficulty for any reasonable value of $p$, including odd and fractional values. We have also shown how the generalized models transition to $\phi^2$ monomials when the tensor-to-scalar ratio $r$ approaches the value $4(1-n_s)$, equivalently $\lambda\rightarrow 0$, where $n_s$ is the spectral index. The resulting models are phenomenologically viable, covering most of  the area preferred by the observations reported by the Planck 2018 collaboration article \cite{Akrami:2018odb}.

\acknowledgments
I would like to thank Prof.~Andrei Linde for informative correspondence and to the anonymous referee for useful advice. We acknowledge financial support from UNAM-PAPIIT,  IN104119, {\it Estudios en gravitaci\'on y cosmolog\'ia}.



\begin{thebibliography}{10}

\bibitem{Akrami:2018odb} 
Y.~Akrami {\it et al.} [Planck Collaboration],
\newblock {Planck 2018 results. X. Constraints on inflation},
\newblock {Astron. Astrophys.}  {\bf 641,} A10 (2020).

\bibitem{Kallosh:2013hoa}
R. Kallosh and A. Linde,
\newblock {Universality Class in Conformal Inflation},
\newblock {J. Cosmol. Astropart. Phys.}, 07 (2013) 002.

\bibitem{Roest:2013fha}
D. Roest,
\newblock {Universality classes of inflation,}
\newblock {J. Cosmol. Astropart. Phys.}, 01 (2014) 007.

\bibitem{Ferrara:2013rsa}
S. Ferrara, R. Kallosh, A. Linde and M. Porrati,
\newblock {Minimal Supergravity Models of Inflation},
\newblock {Phys. Rev. D} {\bf 88}, 085038 (2013).

\bibitem{Kallosh:2013maa}
R. Kallosh and A. Linde,
\newblock {Nonminimal Inflationary Attractors},
\newblock {J. Cosmol. Astropart. Phys.} 10 (2013) 033.

\bibitem{Kallosh:2013daa}
R. Kallosh and A. Linde,
\newblock {Multifield Conformal Cosmological Attractors,}
\newblock {J. Cosmol. Astropart. Phys.} 12 (2013) 006.

\bibitem{Kallosh:2013yoa}
R. Kallosh, A. Linde and D. Roest,
\newblock {Superconformal Inflationary $\alpha$-Attractors},
\newblock {J. High Energy Phys.} {11} (2013) 198.

\bibitem{Kallosh:2013tua}
R. Kallosh, A. Linde and D. Roest,
\newblock {Universal Attractor for Inflation at Strong Coupling},
\newblock {Phys. Rev. Lett},  {\bf 112} 011303 (2014).

\bibitem{Cecotti:2014ipa}
S. Cecotti and R. Kallosh,
\newblock {Cosmological Attractor Models and Higher Curvature Supergravity},
\newblock {J. High Energy Phys.} {05} (2014) 114.

\bibitem{Kallosh:2014rga}
R. Kallosh, A. Linde and D. Roest,
\newblock {Large field inflation and double $\alpha$-attractors},
\newblock {J. High Energy Phys.} {08} (2014) 052.

\bibitem{Kallosh:2014laa}
R. Kallosh, A. Linde and D. Roest,
\newblock {The double attractor behavior of induced inflation},
\newblock {J. High Energy Phys.} {09} (2014) 062.

\bibitem{Galante:2014ifa}
M. Galante, R. Kallosh, A. Linde and D. Roest,
\newblock {Unity of Cosmological Inflation Attractors},
\newblock {Phys. Rev. Lett},  {\bf 114}, 141302 (2015).

\bibitem{Carrasco:2015pla}
J. J. M. Carrasco, R. Kallosh and A. Linde,
\newblock {$\alpha $-Attractors: Planck, LHC and Dark Energy},
\newblock {J. High Energy Phys.} {10} (2015) 147.

\bibitem{Carrasco:2015rva}
J. J. M. Carrasco, R. Kallosh and A. Linde,
\newblock {Cosmological Attractors and Initial Conditions for Inflation},
\newblock {Phys. Rev. D}, {\bf 92}, 063519 (2015).

\bibitem{Carrasco:2015iij}
J. J. M. Carrasco, R. Kallosh and A. Linde,
\newblock {Minimal supergravity inflation},
\newblock {Phys. Rev. D},  {\bf 93}, 061301 (2016).

\bibitem{Kallosh:2016sej}
R. Kallosh, A. Linde, D Roest and T. Wrase,
\newblock {Sneutrino inflation with $\alpha$-attractors},
\newblock {J. Cosmol. Astropart. Phys.} 11 (2016) 046.

\bibitem{Akrami:2017cir}
Y. Akrami, R. Kallosh, A. Linde, and V. Vardanyan,
\newblock {Dark energy, $\alpha$-attractors, and large-scale structure surveys},
\newblock {J. Cosmol. Astropart. Phys.} 06 (2018) 041.

\bibitem{Kallosh:2019hzo}
R. Kallosh and A. Linde,
\newblock {CMB targets after the latest Planck data release},
\newblock {Phys. Rev. D} {\bf 100}, 123523 (2019).

\bibitem{Odintsov:2016vzz}
S. D. Odintsov  and V. K. Oikonomou,
\newblock {Inflationary $\alpha$-attractors from $F(R)$ gravity},
\newblock {Phys. Rev. D}  {\bf 94}, 124026 (2016).

\bibitem{Ueno:2016dim}
Y. Ueno and K. Yamamoto,
\newblock {Constraints on $\alpha$-attractor inflation and reheating},
\newblock {Phys. Rev. D}  {\bf 93}, 083524 (2016).

\bibitem{Kumar:2015mfa}
K. S. Kumar,  J. Marto, P. Vargas Moniz, and S. Das,
\newblock {Non-slow-roll dynamics in $\alpha-$attractors},
\newblock {J. Cosmol. Astropart. Phys.} 04 (2016) 005.

\bibitem{Eshaghi:2016kne}
M. Eshaghi, M. Zarei, N. Riazi and A. Kiasatpour,
\newblock {CMB and reheating constraints to $\alpha$-attractor inflationary models},
\newblock {Phys. Rev. D} {\bf 93}, 123517 (2016).

\bibitem{DiMarco:2017zek}
A. Di Marco, P. Cabella and N. Vittorio,
\newblock {Constraining the general reheating phase in the $\alpha$-attractor inflationary cosmology},
\newblock {Phys. Rev. D} {\bf 95}, 103502 (2017).

\bibitem{Rashidi:2018ois}
N. Rashidi, K. Nozari,
\newblock {$\alpha$-Attractor and reheating in a model with noncanonical scalar fields},
\newblock {Int. J. Mod. Phys. D} {\bf 27}, 1850076 (2018).

\bibitem{Linder:2015qxa}
E.V. Linder,
\newblock {Dark energy from $\alpha$-attractors},
\newblock {Phys. Rev. D} {\bf 91}, 123012 (2015).

\bibitem{Dimopoulos:2017zvq}
K.~Dimopoulos, C.~Owen,
\newblock {Quintessential Inflation with $\alpha$-attractors},
\newblock {J. Cosmol. Astropart. Phys.} 06 (2017) 027.

\bibitem{Dimopoulos:2017tud}
K.~Dimopoulos,  L.~Donaldson Wood and C.~Owen,
\newblock {Instant preheating in quintessential inflation with $\alpha$-attractors},
\newblock {Phys. Rev. D} {\bf 97},  063525 (2018).

\bibitem{Garcia-Garcia:2018hlc}
C. Garcia-Garcia, E.V. Linder, P. Ruiz-Lapuente and M. Zumalacarregui,
\newblock {Dark energy from $\alpha$-attractors: phenomenology and observational constraints},
\newblock {J. Cosmol. Astropart. Phys.} 08 (2018) 022.

\bibitem{Dalianis:2018frf}
I. Dalianis, A. Kehagias and G. Tringas,
\newblock {Primordial black holes from \ensuremath{\alpha}-attractors},
\newblock {J. Cosmol. Astropart. Phys.} 01 (2019) 037.

\bibitem{Cedeno:2019cgr}
F. X. Linares Cede\~no, A. Montiel, J. C. Hidalgo and G. Germ\'an,
\newblock {Bayesian evidence for $\alpha$-attractor dark energy models},
\newblock {J. Cosmol. Astropart. Phys.} 08 (2019) 002.

\bibitem{Shojaee:2020xyr}
R. Shojaee, K. Nozari and F. Darabi,
\newblock {\ensuremath{\alpha}-Attractors and reheating in a non-minimal inflationary model},
\newblock {Int. J. Mod. Phys. D} {\bf 29}, 2050077 (2020).

\bibitem{Odintsov:2020thl}
S. D. Odintsov and  V. K. Oikonomou,
\newblock {Inflationary attractors in $F(R)$ gravity},
\newblock {Phys. Lett. B} {\bf 807}, 135576 (2020).

\bibitem{Akrami:2020zxw}
Y. Akrami, S. Casas, S. Deng, V. Vardanyan,
\newblock {Quintessential $\alpha$-attractor inflation: forecasts for Stage IV galaxy surveys},
\newblock {J. Cosmol. Astropart. Phys.} 04 (2021) 006.

\bibitem{Shojaee:2021zmf}
R. Shojaee, K. Nozari and F. Darabi.
\newblock {$\alpha$-Attractors and reheating in a class of Galileon inflation},
\newblock {Int. J. Mod. Phys. D} {\bf 30}, 2150036 (2021).

\bibitem{Pal:2009sd} 
B.~K.~Pal, S.~Pal and B.~Basu,
\newblock {Mutated Hilltop Inflation : A Natural Choice for Early Universe},
\newblock {J. Cosmol. Astropart. Phys.} 01 (2010) 029.

\bibitem{Pal:2010eb} 
B.~K.~Pal, S.~Pal and B.~Basu,
\newblock {A semi-analytical approach to perturbations in mutated hilltop inflation},
\newblock {Int.\ J.\ Mod.\ Phys.} D {\bf 21}, 1250017 (2012).

\bibitem{German:2020eyq} 
G.~Germ\'an,
\newblock {Constraints for the running index independent of the parameters of the model},
{Int. J. Mod. Phys. D}  {\bf 30}, 2150038 (2021).

\bibitem{German:2020cbw} 
G.~Germ\'an,
\newblock {Constraints from reheating},
arXiv:2010. 09795.

\bibitem{German:2021tqs} 
G.~Germ\'an,
\newblock {On the $\alpha$-attractor $T$ models},
\newblock {J. Cosmol. Astropart. Phys.} 09 (2021) 017.


 \end{thebibliography}
\end{document}